\documentclass[12pt]{article}
\usepackage{amsfonts,amsmath,amssymb,mathrsfs}

\title{On the Role of Density Matrices in Bohmian
     Mechanics}
\author{
   Detlef D\"urr\footnote{Mathematisches Institut der Universit\"{a}t
         M\"{u}nchen, Theresienstra{\ss}e 39, 80333 M\"{u}nchen, Germany.
         E-mail: duerr@mathematik.uni-muenchen.de},\
   Sheldon Goldstein\footnote{Departments of Mathematics, Physics, and
         Philosophy, Hill
         Center, Rutgers, The State University of New Jersey, 110
         Frelinghuysen Road, Piscataway, NJ 08854-8019, USA. E-mail:
         oldstein@math.rutgers.edu},\\
   Roderich Tumulka\footnote{Dipartimento di Fisica and INFN sezione di
         Genova, Universit\`a di Genova, Via Dodecaneso 33,
         16146 Genova, Italy.
         E-mail: tumulka@mathematik.uni-muenchen.de},\ and
   Nino Zangh\`\i\footnote{Dipartimento di Fisica and INFN sezione di
         Genova, Universit\`a di Genova, Via Dodecaneso 33,
         16146 Genova, Italy. E-mail: zanghi@ge.infn.it}
}
\date{November 19, 2003}

\newcommand{\CCC}{\mathbb{C}} 
\newcommand{\RRR}{\mathbb{R}} 
\newcommand{\I}{i} 
\newcommand{\tr}{\mathrm{tr}} 
\newcommand{\Laplace}{\Delta} 

\renewcommand{\Im}{\mathrm{Im}} 

\newcommand{\vq}{{\boldsymbol q}}
\newcommand{\vQ}{{\boldsymbol Q}}

\newcommand{\Hilbert}{{\cal H}}
\renewcommand{\sp}[2]{\langle #1 | #2 \rangle}
\newcommand{\conf}{\mathcal{Q}} 
\renewcommand{\div}{\,\mathrm{div}\,} 
\newcommand{\prob}{\mathrm{Prob}}
\newcommand{\pov}[1]{ \hat{P}_{\hat{q}_{#1}} }
\newcommand{\experiment}{\mathcal{E}}
\newcommand{\valuespace}{\mathcal{V}}
\newcommand{\cond}{\mathrm{cond}} 
\newcommand{\stat}{\mathrm{stat}} 
\newcommand{\red}{\mathrm{red}}   
\newcommand{\comb}{\mathrm{comb}} 
\newcommand{\fund}{\mathrm{fund}} 

\newcommand{\sphere}{\mathscr{S}(\Hilbert)}
\newcommand{\SSS}{\mathscr{S}}
\newcommand{\nnn}{\mathcal{N}}
\begin{document}\maketitle
\begin{abstract}
   It is well known that density matrices can be used in quantum
   mechanics to represent the information available to an observer
   about either a system with a random wave function (``statistical
   mixture'') or a system that is entangled with another system
   (``reduced density matrix''). We point out another role, previously
   unnoticed in the literature, that a density matrix can play: it can
   be the ``conditional density matrix,'' conditional on the
   configuration of the environment. A precise definition can be given
   in the context of Bohmian mechanics, whereas orthodox quantum
   mechanics is too vague to allow a sharp definition, except perhaps
   in special cases. In contrast to statistical and reduced density
   matrices, forming the conditional density matrix involves no
   averaging. In Bohmian mechanics with spin, the conditional density
   matrix replaces the notion of conditional wave function, as the
   object with the same dynamical significance as the wave function of
   a Bohmian system.

   \medskip

\noindent PACS number: 03.65.Ta (foundations of quantum mechanics)

\end{abstract}

\section{Introduction}
We wish to dedicate this work to the memory of Jim Cushing, our
friend, coworker and colleague.

In this paper we shall be concerned with the following claim: Once we
deal with particles with spin in Bohmian mechanics, we are more or
less obliged to regard the quantum state of any system (except the
universe) as given by a density matrix, which then has precisely the
same dynamical significance as the wave function. The aim of this
paper is to elaborate on this statement, as it is far {}from obvious
in what sense a density matrix could represent the dynamical state of
a Bohmian system. In fact, our statement is in sharp contrast with
that of Bell \cite{Belldensity}:
\begin{quotation}
   \noindent So in the de~Broglie--Bohm theory a fundamental
   significance is given to the wave function, and it cannot be
   transferred to the density matrix.
\end{quotation}
Although this is correct for spin 0 particles, the situation changes
as soon as we consider spin or any other internal degree of freedom.
To appreciate this point, it is essential to distinguish between
different roles that density matrices can play in Bohmian mechanics
(or, for that matter, in other versions of quantum mechanics). In one
of these roles, the density matrix is of a purely epistemic character,
i.e., it expresses ignorance, whereas in another role, a role that has
as yet not been discussed in the literature and of which Bell was
obviously not aware, a density matrix is of direct significance to the
Bohmian particle motion, as the ``conditional density matrix.''

We distinguish in this paper five roles of density matrices:
 the statistical, reduced, combined (reduced
statistical), conditional, and fundamental density matrix. We explain
the relations between them and their relevance to the particle motion.
We explain in particular the new notion of conditional density matrix
and its relevance to Bohmian mechanics.

A particular consequence of our discussion is that the same system
can, at one and  the same time, have a conditional
density matrix and, say, a different reduced density
matrix.  Thus, when speaking about ``the'' density
matrix of a system, it is necessary to specify whether one refers to
the reduced or the conditional density matrix. This is new: among the
traditional types of density matrices, it is always clear (except for
the ambiguity in some cases as to whether one should consider
collapsed or uncollapsed wave functions) which type of density matrix
is relevant to a given system, and what this density matrix is---so
that it is possible to speak of \emph{the} density matrix of the
system. The fact that a system can have two different density matrices
at the same time is why we have to focus on the \emph{role} that a
density matrix plays for the theoretical treatment of a system, since
that is the only way to understand how more than one density matrix
can be relevant to the same system.

\section{Bohmian Mechanics}\label{sec:BM}

We begin by briefly recalling Bohmian mechanics. It is a theory of
point particles moving in physical space $\RRR^{3}$. For the sake of
concreteness, consider a universe of $N$ nonrelativistic particles
whose positions we denote by $\vQ_1(t), \ldots, \vQ_N(t)$. They move
according to Bohm's equation of motion,
\begin{equation}\label{Bohm}
   \frac{d\vQ_j}{dt} = \frac{\hbar}{m_j} \Im \frac{\psi^* \nabla_j \psi}
   {\psi^* \psi} (\vQ_1, \ldots, \vQ_N)
\end{equation}
where $m_j$ is the mass of particle $j$, $\psi: \RRR^{3N} \to \CCC^k$
is the wave function, and $\psi^* \psi$ denotes the scalar product in
$\CCC^k$. In the case $k=1$ (spin 0), \eqref{Bohm} simplifies to
\begin{equation}\label{complexBohm}
   \frac{d\vQ_j}{dt} = \frac{\hbar}{m_j} \Im \frac{\nabla_j \psi}
   {\psi} (\vQ_1, \ldots, \vQ_N) \,.
\end{equation}
$\psi$ evolves according to the Schr\"odinger equation
\begin{equation}\label{Schroedinger}
   \I \hbar \frac{\partial \psi}{\partial t} = - \sum_{j=1}^N
   \frac{\hbar^2}{2m_j} \Laplace_j \psi + V\psi =: \hat{H} \psi
\end{equation}
where the potential $V$ may take values in the $k \times k$
Hermitian matrices. The configuration $Q(t) =
(\vQ_1(t), \ldots, \vQ_N(t))$ is random and $|\psi(t)|^2$-distributed
at every time $t$,
\begin{equation}\label{rhopsi}
   \prob (Q(t) \in dq) = |\psi(q,t)|^2 dq\,.
\end{equation}
This is possible because of an equivariance property of \eqref{Bohm}
and \eqref{Schroedinger}: if \eqref{rhopsi} holds at $t=0$ then it
also holds at every other time. This follows {}from the following
continuity equation, a consequence of \eqref{Schroedinger}:
\begin{equation}
\frac{\partial |\psi|^2}{\partial t} = - \div (|\psi|^2 v)
\end{equation}
where $v$ is the velocity field, i.e., the (time-dependent) vector
field on $\RRR^{3N}$ whose $j$-th component is the right hand side of
\eqref{Bohm}. We remark that the state at time $t$ of a Bohmian
universe is described by the pair $(Q(t),\psi(t))$.

What we describe in this paper about conditional density matrices
applies not only to conventional nonrelativistic Bohmian mechanics as
just described, but also to Bohmian mechanics on curved manifolds
\cite{jamesthesis,identical}, to Bohm's trajectories for Dirac wave
functions (see \cite[p.~272]{BohmHiley} and \cite{HBD}), to the photon
trajectories of \cite{photon}, to the jump processes of \cite{crea2},
and, in a sense that we will explain more fully in
Section~\ref{sec:region}, also to theories with a variable number of
particles \cite{crlet,BellBeables,crea1,crea2}.

\section{Three Density Matrices}\label{sec:3}

If $\Hilbert$ denotes the Hilbert space of a system $S$, a density
matrix for $S$ is a positive, (bounded) self-adjoint operator
$\hat{W}: \Hilbert \to \Hilbert$ with $\tr \, \hat{W} =1$. If, as in
Bohmian mechanics, $\Hilbert$ is a space of wave functions on a
configuration space $\conf$, $\Hilbert = L^2 (\conf, \CCC^k)$, then a
density matrix can also be viewed as a function $W: \conf \times \conf
\to \mathrm{End}(\CCC^k)$ (where $\mathrm{End}(\CCC^k)$ denotes the
space of linear mappings (endomorphisms) $\CCC^k \to \CCC^k$).  The
translating relations between the two views, operator on $\Hilbert$
and function on $\conf \times \conf$, are
\begin{subequations}
\begin{align}
   \big(\hat{W} \psi\big)^s (q) &= \int\limits_\conf dq' \sum_{s'}
   W^s_{s'} (q,q') \, \psi^{s'}(q') \text{ and}\\
   W^s_{s'} (q,q') &= \sp{q,s}{\hat{W}|q',s'}
\end{align}
\end{subequations}
where $s$ and $s'$ index the standard basis of $\CCC^k$. The function
$W$ has the properties
\begin{subequations}\label{Wproperties}
\begin{align}
   &W(q',q) = W^*(q,q') \label{sa} \\
   0 \leq \int\limits_\conf dq \int\limits_\conf dq' \sum_{s,s'}
   &\psi^*_s (q) \, W^s_{s'}(q,q') \, \psi^{s'}(q') < \infty \quad
   \forall \psi \in \Hilbert \label{positive} \\
   &\int\limits_\conf dq \, \tr_{\CCC^k} \, W(q,q) =
   1,\label{normalized}
\end{align}
\end{subequations}
where $W^*$ denotes the adjoint endomorphism in $\CCC^k$, whose matrix
is the conjugate transposed. Conversely, the properties
\eqref{Wproperties} are sufficient for $W$ to define a density matrix
$\hat{W}$. A particular consequence of \eqref{sa} is that on the
diagonal of $\conf \times \conf$, $W(q,q)$ is a Hermitian endomorphism
(and thus $\tr_{\CCC^k} \, W(q,q) \in \RRR$), and a particular
consequence of \eqref{positive} is that
\begin{equation}\label{posdiagonal}
   \tr_{\CCC^k} \, W(q,q) \geq 0 \quad \forall q \in \conf.
\end{equation}

There are four ways in which density matrices can
arise {}from Bohmian or quantum mechanics. Three of them are well
known; we briefly recall them anyway.
\begin{enumerate}
\item First, by statistical mixture. Suppose the wave function $\psi$
   of a system is random with probability distribution $\mu(d\psi)$ on
   the unit sphere $\sphere$ of the Hilbert space
   $\Hilbert$.  The associated \emph{statistical density matrix} is
\begin{subequations}
\begin{align}
   \hat{W}_\stat &= \int\limits_{\sphere} \mu(d\psi) \, |\psi \rangle
   \langle \psi| \intertext{respectively} {W_\stat}^s_{s'}(q,q') &=
   \int\limits_{\sphere} \mu(d\psi) \, \psi^s(q) \, \psi^*_{s'}(q') \,.
\end{align}
\end{subequations}
This density matrix was first considered in \cite{Neumann27}. Note
that different distributions $\mu$ may lead to the same density
matrix. (For example, the density matrix
$\frac{1}{k} I$ on the finite-dimensional Hilbert space $\CCC^k$
arises from the discrete uniform distribution over the vectors of any
orthonormal basis in $\CCC^k$, as well as from the continuous uniform
distribution over the unit sphere $\SSS(\CCC^k)$.)  The significance
of $\hat{W}_\stat$ lies in the fact that the distribution of the
random outcome $Z$ of an experiment performed on the system depends on
$\mu$ only trough $\hat{W}_\stat$; i.e., different
$\mu$'s leading to the same density matrix also lead to the same
statistics of outcomes. More precisely, when the experiment ``measures
the observable'' $\hat{A}$, the probability of obtaining an outcome
$Z$ in the set $B \subseteq \RRR$ is
\begin{equation}\label{prob1}
   \prob (Z \in B) = \tr \big(\hat{W}_\stat \hat{P}_{\hat{A}}(B) \big)
\end{equation}
where $\hat{P}_{\hat{A}}$ is the projection-valued measure (PVM) on
the real line given by the spectral decomposition of the self-adjoint
operator $\hat{A}$.\footnote{We remind the reader that in Bohmian
   mechanics such an experiment need not \emph{measure} anything in the
   literal sense of the word \cite{survey,op}. We also note that
   \eqref{prob1} holds not only for ``measurements of observables,''
   but for arbitrary experiments $\experiment$ with results in the
   value space $\valuespace$: with every $\experiment$ is associated a
   positive-operator-valued measure (POVM) $\hat{P}_{\experiment}$
   \cite{Davies,op} such that the probability of obtaining {}from
   $\experiment$ an outcome in the set $B \subseteq \valuespace$ is
   $\tr \big(\hat{W}_\stat \hat{P}_{\experiment}(B) \big)$.}  This
follows by averaging, according to $\mu$, of the probability that the
result is in $B$ given that the state vector of the
system is $\psi$, which is (in both standard quantum mechanics and
Bohmian mechanics) $\sp{\psi}{\hat{P}_{\hat{A}}(B)| \psi}$. A
particular consequence of \eqref{prob1} is that the outcomes of
position measurements are distributed according to the density
\begin{equation}\label{prob5}
   \rho(q) = \tr_{\CCC^k} W_\stat(q,q)
\end{equation}
on configuration space $\conf$.

{}From Schr\"odinger's equation \eqref{Schroedinger} for $\psi$, one
obtains an evolution law \cite{Neumann} for $\hat{W}_\stat$:
\begin{subequations}\label{vonNeumann}
\begin{align}
   \I \hbar \frac{\partial \hat{W}_\stat}{\partial t} &= [\hat{H},
   \hat{W}_\stat] \intertext{respectively} \I \hbar \frac{\partial
     W_\stat(q,q')}{\partial t} &= \hat{H}_{q} W_\stat(q,q') -
   \hat{H}_{q'} W_\stat(q,q')
\end{align}
\end{subequations}
where $\hat{H}_q$ means that the Hamiltonian $\hat{H}$ acts on the
variable $q$, and $[\;, \,]$ denotes the commutator.  We remark that
$\hat{W}_\stat$ is ``pure,'' i.e., a projection to a 1-dimensional
subspace, if and only if $\mu$ is concentrated on that subspace.

\item The second situation in which a density matrix is relevant
   involves a system $S_1$ that is entangled with another system $S_2$.
   In this case, the composite system $S_1 \cup S_2$ possesses a wave
   function $\Psi^{s_1 s_2}(q_1,q_2)$ or $\Psi \in \Hilbert_1 \otimes
   \Hilbert_2$, but no wave function is associated with $S_1$ alone.
   However, the following \emph{reduced density matrix} can be
   associated with $S_1$:
\begin{subequations}
\begin{align}
   \hat{W}_\red &= \tr_2 \, |\Psi \rangle \langle \Psi|
   \intertext{respectively} {W_\red}^{s_1}_{s_1'} (q_1,q_1') &=
   \int\limits_{\conf_2} dq_2 \sum_{s_2} \Psi^{s_1 s_2}(q_1,q_2) \,
   \Psi^*_{s_1' s_2} (q_1',q_2)
\end{align}
\end{subequations}
where $\tr_2$ denotes the partial trace over $\Hilbert_2$. This kind
of density matrix was first considered in \cite{redu}. Note that
$\hat{W}_\red$ is an operator on $\Hilbert_1$. Like $\hat{W}_\stat$,
$\hat{W}_\red$ possesses significance in terms of probability
distributions: if one ``measures'' $\hat{A}$ on $S_1$ alone, then the
probability of obtaining a result $Z$ in the set $B \subseteq \RRR$ is
\begin{equation}\label{prob2}
   \prob (Z \in B) = \tr \big(\hat{W}_\red \hat{P}_{\hat{A}}(B) \big)
\end{equation}
where the trace is, of course, taken in $\Hilbert_1$. This equation
follows {}from the fact that the observable on $\Hilbert_1 \otimes
\Hilbert_2$ that corresponds to this experiment, as an experiment on
$S_1 \cup S_2$, is $\hat{A} \otimes \hat{1}$, so that the probability
for $Z \in B$ is $\sp{\Psi}{\hat{P}_{\hat{A}}(B) \otimes \hat{1}
   |\Psi}$, which equals \eqref{prob2}.

If $S_1$ and $S_2$ are decoupled, i.e., if $\hat{H} = \hat{H}_1
\otimes \hat{1} + \hat{1} \otimes \hat{H}_2$, the reduced density
matrix evolves in the same way as statistical density matrices do,
governed by $\hat{H}_1$:
\begin{subequations}\label{vonNeumannred}
\begin{align}
   \I \hbar \frac{\partial \hat{W}_\red}{\partial t} &= [\hat{H}_1,
   \hat{W}_\red] \intertext{respectively} \I \hbar \frac{\partial
     W_\red(q,q')}{\partial t} &= \hat{H}_{1q} W_\red(q,q') -
   \hat{H}_{1q'} W_\red(q,q')\,.
\end{align}
\end{subequations}
In case $S_1$ and $S_2$ are coupled, $W_\red$ does not have an
autonomous dynamics, i.e., its evolution depends on the $\Psi$ {}from
which it arises.  We remark that $\hat{W}_\red$ is ``pure'' if and
only if $S_1$ and $S_2$ are disentangled, $\Psi^{s_1s_2}(q_1,q_2) =
\psi_1^{s_1}(q_1) \, \psi_2^{s_2}(q_2)$.

\item The third possibility is the combination of the first and the
   second types of density matrices: the reduced density matrix of a
   statistical mixture. Suppose the wave function $\Psi$ of the system
   $S_1 \cup S_2$ is random with distribution $\mu$ on $\SSS(\Hilbert_1
   \otimes \Hilbert_2)$. Then define the \emph{combined density matrix}
   by
\begin{subequations}
\begin{align}
   \hat{W}_\comb &= \int\limits_{\SSS(\Hilbert_1 \otimes \Hilbert_2)}
   \mu(d\Psi) \, \tr_2 \, |\Psi \rangle \langle \Psi|
   \intertext{respectively} {W_\comb}^{s_1}_{s_1'} (q_1,q_1') &=
   \int\limits_{\SSS(\Hilbert_1 \otimes \Hilbert_2)} \mu(d\Psi)
   \int\limits_{\conf_2} dq_2 \sum_{s_2} \Psi^{s_1 s_2} (q_1,q_2) \,
   \Psi^*_{s_1' s_2} (q_1', q_2) \,.
\end{align}
\end{subequations}
This kind of density matrix was first considered in
\cite[p.~424]{Neumann}.  $\hat{W}_\comb$ can be obtained either by
averaging the reduced density matrix associated with the random state
$\Psi$, or by reducing, i.e., taking the partial trace of, the
statistical density matrix on $\Hilbert_1 \otimes \Hilbert_2$
associated with $\mu$.  Again, the probability that the result $Z$ of
an experiment on $S_1$ ``measuring'' $\hat{A}$ lies in the set $B
\subseteq \RRR$ is
\begin{equation}\label{prob3}
   \prob(Z \in B) = \tr \big( \hat{W}_\comb \hat{P}_{\hat{A}}(B) \big) 
\,.
\end{equation}
This follows either {}from averaging \eqref{prob2} over $\mu$ or
{}from applying \eqref{prob1} to $\hat{A} \otimes \hat{1}$.

Like the reduced density matrix, $\hat{W}_\comb$ follows the unitary
evolution governed by $\hat{H}_1$ whenever that makes sense, i.e.,
whenever $S_1$ and $S_2$ are decoupled.  $\hat{W}_\comb$ is pure if
and only if $\mu$ is concentrated on the subspace $\CCC \psi_1 \otimes
\Hilbert_2$ for some $\psi_1 \in \Hilbert_1$.
\end{enumerate}

\section{A Fourth Density Matrix}\label{sec:4}

We now turn to the fourth, novel, kind of density matrix: the
conditional density matrix. It also involves a system $S_1$ that is
entangled with $S_2$, and it is related to the notion of conditional
wave function \cite{qe} which we recall first. For the sake of
definiteness, we take $S_2$ to be the environment
of $S_1$, i.e., the rest of the universe.

In Bohmian mechanics for spin 0 particles, more precisely in Bohmian
mechanics with complex-valued wave functions, the \emph{conditional
   wave function} of $S_1$ is obtained {}from the wave function
$\Psi(q_1,q_2)$ of $S_1 \cup S_2$ by inserting the actual
configuration $Q_2$ of $S_2$,
\begin{subequations}
\begin{align}
  \label{psiconddef}
  \psi_\cond (q_1) &= \frac{1}{\sqrt{\nnn}} \Psi(q_1,Q_2)\\
  \text{where } \nnn &= \int\limits_{\conf_1} dq_1 \, |\Psi(q_1,Q_2)|^2
\end{align}
\end{subequations}
is a normalizing factor ensuring that $\int |\psi_\cond|^2 =1$.
$\psi_\cond$ can be viewed as the wave function of $S_1$ alone. It
does not, in general, evolve according to a Schr\"odinger equation
\eqref{Schroedinger}, indeed it does not have an autonomous dynamics
at all.\footnote{The conditional wave function at time $t=0$ need not
   determine the conditional wave function at later times. As an
   example, consider two situations with the same $\Psi$, the same
   $Q_2(0)$ and different $Q_1(0)$: since $\psi_\cond$ does not depend
   on $Q_1$, it will be the same in the two situations at $t=0$, but
   since the motion of $Q_2$ typically depends on $Q_1$, the two
   situations will typically have different $Q_2$'s at later times, and
   thus typically different $\psi_\cond$'s.} In
fact, in appropriate situations the evolution of
$\psi_\cond$ leads to collapse, in the usual textbook manner, which
seems quite appropriate for the wave function of a subsystem.
$\psi_\cond$ shares the following basic properties with the wave
function $\psi$ in Bohmian mechanics:

\begin{itemize}
\item The conditional distribution of $Q_1$ given $Q_2$ is
   $|\psi_\cond|^2$.  More precisely, we have the
   following formula for the conditional probability:
\begin{equation}\label{condprob}
   \prob ( Q_1 \in dq_1 | Q_2 ) = |\psi_\cond(q_1) |^2 dq_1\,,
\end{equation}
which resembles the formula \eqref{rhopsi} for the probability in
terms of the wave function.  \eqref{condprob} follows {}from the fact
that the pair $(Q_1,Q_2)$ is $|\Psi|^2$ distributed.

\item The motion of $Q_1$ can be computed {}from $\psi_\cond$
   according to
\begin{equation}\label{condBohm}
   \frac{d\vQ_{1j}}{dt} = \frac{\hbar}{m_{1j}} \Im \frac{\nabla_{1j}
   \psi_\cond} {\psi_\cond}(\vQ_{11}, \ldots, \vQ_{1N_1})\,,
\end{equation}
which is the same formula as \eqref{complexBohm} for the velocity in
terms of the wave function.
\end{itemize}

An analogous conditional wave function cannot be formed, however, when
the particles of $S_2$ have spin or any other internal degree of
freedom entailing that the wave function has several complex
components. The reason is that $\psi_\cond$ as defined in
\eqref{psiconddef} would have too many components, i.e., more spin
indices than appropriate for a wave function of $S_1$ alone. In
particular, $\psi_\cond$ would not be an element of $\Hilbert_1$.

We propose to consider instead the \emph{conditional density matrix},
which is obtained {}from $\Psi(q_1,q_2) \, \Psi^* (q_1', q_2')$ by
inserting the actual configuration $Q_2$ of $S_2$ for both $q_2$ and
$q_2'$, and contracting over the spin index belonging to $S_2$:
\begin{equation}\label{Wconddef}
   {W_\cond}^{s_1}_{s_1'}(q_1,q_1') = \frac{1}{\nnn} \sum_{s_2} 
\Psi^{s_1 s_2}
   (q_1,Q_2) \, \Psi^*_{s_1' s_2} (q_1',Q_2)
\end{equation}
with normalizing factor\footnote{One can show that for almost every
   configuration $Q = (Q_1,Q_2)$ (almost every with respect to the
   $|\Psi|^2$ distribution), $\nnn$ will be neither zero nor infinite.}
\begin{equation}\label{ndef}
   \nnn = \int\limits_{\conf_1} dq_1 \sum_{s_1 s_2}
   \Psi^{s_1 s_2}(q_1,Q_2) \, \Psi^*_{s_1 s_2} (q_1, Q_2) \,.
\end{equation}
One easily checks that $W_\cond$ satisfies \eqref{Wproperties} and
thus is a density matrix.\footnote{The only step that may not be
   obvious is the finiteness part of \eqref{positive}, which follows
   {}from the fact that $\nnn<\infty$ so that for any fixed value of
   $s_2$, $\Psi^{s_1s_2}(q_1,Q_2)$ as a function of $s_1$ and $q_1$
   lies in $L^2(\conf_1,\CCC^{k_1})$; thus the scalar product with any
   $\psi \in L^2(\conf_1, \CCC^{k_1})$ is finite.} The expression for
the corresponding operator $\hat{W}_\cond$ reads
\begin{equation}\label{Wopdef}
   \hat{W}_\cond = \frac{\tr_2 \big( |\Psi \rangle \langle \Psi |
   \hat{1} \otimes \pov{2}(dq_2) \big)}{\tr \big( |\Psi \rangle \langle
\Psi |
   \hat{1} \otimes \pov{2}(dq_2) \big)} (q_2 =Q_2)
\end{equation}
where $\pov{2}$ is the projection-valued measure on $\conf_2$ defined
by the joint spectral decomposition of all position operators of
$S_2$, and the fraction is a Radon--Nikod\'ym derivative of an
operator-valued measure on $\conf_2$ with respect to a real-valued
measure on $\conf_2$, and thus an operator-valued function on
$\conf_2$, into which we insert $Q_2$.

We remark that $\hat{W}_\cond$ is pure if and only
if $\Psi(q_1,Q_2)$ as an element of $L^{2}(\mathcal{Q}_{1},\CCC^{k_1})
\otimes \CCC^{k_2}$ is a tensor product, $\Psi^{s_1 s_2}(q_1, Q_{2}) =
\psi_1^{s_1} (q_1) \, \psi_2^{s_2}$.  In particular, $\hat{W}_\cond$
is pure if $\Psi$ is complex valued.

The conditional density matrix has the following properties analogous
to those of the conditional wave function:

\begin{itemize}
\item The conditional distribution of $Q_1$ given
   $Q_2$ can be computed {}from $W_\cond$ by taking
   the trace on the diagonal. More precisely, we have the following
   formula for the conditional probability:
\begin{equation}\label{Wcondprob}
   \prob ( Q_1 \in dq_1 | Q_2 ) = \tr_{\CCC^{k_1}} \, W_\cond (q_1,q_1)
   \, dq_1\,.
\end{equation}
This follows {}from the fact that the pair $(Q_1,Q_2)$ is $|\Psi|^2$
distributed. Note that the right hand side is the usual expression
\eqref{prob5} for the probability distribution on configuration
space when a system is described by a density
matrix.

\item The motion of $Q_1$ can be computed {}from $W_\cond$ according
   to
\begin{equation}\label{WcondBohm}
   \frac{d\vQ_{1j}}{dt} = \frac{\hbar}{m_{1j}} \Im 
\frac{\nabla_{\vq_{1j}}
   \tr_{\CCC^{k_1}} \, W_\cond(q_1,q_1')} {\tr_{\CCC^{k_1}} \,
   W_\cond(q_1,q_1')}(q_1 = q_1' = Q_1)\,.
\end{equation}
\end{itemize}

To be able to appreciate \eqref{WcondBohm}, we have to consider a
fifth type of density matrix.

\section{A Fifth Density Matrix}\label{sec:5}

A density matrix is relevant in yet another way: in a modified version
of Bohmian mechanics in which the particles are guided not by a wave
function but by a density matrix. Let us call this $W$-Bohmian
mechanics. Whereas in the conventional version of Bohmian mechanics
the wave function (of the universe) is something
real, as an objective component of the state of the
universe at a given time, in $W$-Bohmian mechanics instead of a wave
function (of the universe) we may have only a density matrix.  This
density matrix does not arise in any way {}from an analysis of the
theory, but is built into the fundamental postulates of $W$-Bohmian
mechanics. It is a \emph{fundamental density matrix}, $W_\fund$, in
contrast to the four other density matrices we have discussed,
 which were derived objects, derived {}from $\psi$
and $Q$. Like the conditional density matrix, $W_\fund$ has not been
considered previously in the literature. The state at time $t$ of a
$W$-Bohmian universe is given by the pair $(Q(t),W_\fund(t))$, and it
evolves according to
\begin{equation}\label{WBohm}
   \frac{d\vQ_j}{dt} = \frac{\hbar}{m_j} \Im \frac{\nabla_{\vq_j}
   \tr_{\CCC^k} \, W_\fund(q,q')} {\tr_{\CCC^k} \, W_\fund(q,q')} 
(q=q'=Q)
\end{equation}
as the equation of motion for $Q$, and
\begin{subequations}\label{WSchroedinger}
\begin{align}
   \I \hbar \frac{\partial \hat{W}_\fund}{\partial t} &= [\hat{H},
   \hat{W}_\fund] \intertext{respectively} \I \hbar \frac{\partial
     W_\fund(q,q')}{\partial t} &= \hat{H}_{q} W_\fund(q,q') -
   \hat{H}_{q'} W_\fund(q,q')
\end{align}
\end{subequations}
for $\hat{W}_\fund$, respectively $W_\fund(q,q')$. 
Note that equations \eqref{WSchroedinger} are the same as
\eqref{vonNeumann} and \eqref{vonNeumannred}.  \eqref{WBohm} was first
written down by Bell \cite{Belldensity} for the purpose of contrasting
it  with the implications of Bohm's equation of
motion \eqref{Bohm} for a system with a random wave function, hence
described by $\hat{W}_\stat$.

The configuration $Q(t)$ is random with distribution given by the
trace of the diagonal of $W_\fund(t)$, i.e.,
\begin{equation}\label{Wprob}
   \prob (Q(t) \in dq) = \tr_{\CCC^k} \, W_\fund(q,q,t) \, dq.
\end{equation}
This is possible because of the following equivariance theorem: if
\eqref{Wprob} holds at $t=0$ then it also holds at every other time.
To see this, note that \eqref{WSchroedinger} implies that
\begin{equation}
  \frac{\partial \tr_{\CCC^k} \, W_\fund(q,q)}{\partial t} = - \div
  (\tr_{\CCC^k} \, W_\fund(q,q) \, v)
\end{equation}
where $v$ is the velocity field, i.e., the (time-dependent) vector
field on $\conf$ whose $j$-th component is the right hand side of
\eqref{WBohm}.

\section{Discussion}\label{sec:discussion}

Bohmian mechanics, as described in Section \ref{sec:BM}, is a special
case of $W$-Bohmian mechanics: if $W_\fund$ is pure, i.e., if it
arises {}from a wave function $\psi$ via 
\begin{equation}\label{Wpsi}
   {W_\fund}^s_{s'}(q,q') = \psi^s(q) \, \psi^*_{s'}(q')\,,
\end{equation}
then the equation of motion \eqref{WBohm} reduces to Bohm's equation
of motion \eqref{Bohm}, the probability law \eqref{Wprob} reduces to
the $|\psi|^2$ law \eqref{rhopsi}, and the evolution
\eqref{WSchroedinger} entails that $W_\fund$ remains pure and arises
{}from a wave function that evolves according to the Schr\"odinger
equation \eqref{Schroedinger}.

Conversely, the equations \eqref{WBohm} and \eqref{Wprob} of
$W$-Bohmian mechanics arise for the behavior of subsystems {}from
Bohmian mechanics for systems of many particles with spin: The motion
of the particles of subsystem $S_1$ is governed according to
\eqref{WcondBohm} by a density matrix, $\hat{W}_\cond$, in the same
way as in $W$-Bohmian mechanics the motion of particles is governed
according to \eqref{WBohm} by a density matrix, $\hat{W}_\fund$.  In
addition to the velocities, also the probabilities \eqref{Wcondprob}
are determined by a density matrix in the same way as in $W$-Bohmian
mechanics \eqref{Wprob}. Thus, even were the universe as a whole
governed by Bohmian mechanics, for most subsystems the state would be
described by a density matrix, $W_\cond$, with the velocities and
probabilities of the subsystem governed by the equations of
$W$-Bohmian mechanics for $W_\cond$.  In this sense, $W$-Bohmian
mechanics is the theory relevant to most systems in a Bohmian
universe.  (More precisely, this holds for all those systems for which
$\hat{W}_\cond$ is not pure.)

A big difference, however, between the dynamics of a subsystem and
$W$-Bohmian mechanics lies in the fact that, 
unlike the fundamental density matrix, see \eqref{WSchroedinger}, the
conditional density matrix need not evolve unitarily. Nevertheless,
there are special situations in which $W_\cond$ does evolve unitarily,
at least as a good approximation. This happens trivially when $S_1$
and $S_2$ are disentangled, $\Psi(q_1,q_2) = \psi_1(q_1) \otimes
\psi_2(q_2)$, and decoupled (so that they stay disentangled). It also
happens when (and for as long as) $S_1$  and $S_2$
are decoupled and
\begin{equation}
  \Psi(q_1,q_2) = \psi_1(q_1) \otimes \psi_2(q_2)
   + \Psi^\perp(q_1,q_2),
\end{equation}
i.e.,
\begin{equation}\label{eff1}
   \Psi^{s_1s_2}(q_1,q_2) = \psi_1^{s_1}(q_1) \, \psi_2^{s_2}(q_2)
   + (\Psi^\perp)^{s_1s_2}(q_1,q_2),
\end{equation}
where $\psi_2$ and $\Psi^\perp$ have disjoint $q_2$-supports and $Q_2
\in \mathrm{support} \, \psi_2$. Such a situation often occurs after a
measurement, and indeed allows us to regard $\psi_1$ as the
(effective) wave function of $S_1$, obeying Schr\"odinger's equation
\eqref{Schroedinger}. For $\mathrm{spin}$ $0$, \eqref{eff1}
characterizes the situation in which we can expect the conditional
wave function to evolve unitarily; thus, the conditional density
matrix evolves unitarily in all situations in which the conditional
wave function would for $\mathrm{spin}$ $0$.  We obtain another case
of unitarily evolving $W_\cond$ by replacing \eqref{eff1} by
\begin{equation}\label{eff2}
   \Psi^{s_1s_2}(q_1,q_2) = \psi_1^{s_1s_2}(q_1) \, \psi_2(q_2)
   + (\Psi^\perp)^{s_1s_2}(q_1,q_2),
\end{equation}
with a complex-valued $\psi_2$, and assuming in addition that the
Hamiltonian $\hat{H}_{2}$ for $S_2$  involves no
interaction between spin and configurational degrees of freedom.

For example, consider an EPR--Bohm--Bell pair of spin 1/2 particles,
each headed towards its Stern--Gerlach magnet, with $q_{1}$ and
$q_{2}$ the positions of the particles. Suppose both magnets are
oriented so as to measure $\sigma_{z}$ and that the geometry is such
that particle 1 completely passes its SG magnet before particle 2
reaches its SG magnet. Initially the spin state is the singlet state,
depending on neither $q_{1}$ nor $q_{2}$, and we may assume as well
that the configuration space wave packet is initially of product form
$\psi_{1}(q_{1})\psi_{2}(q_{2})$. Then the initial wave function is of
the form \eqref{eff2} with $\Psi^\perp = 0$ and (regarding the
possible values of $s_{i}$ as $\pm 1$)
\begin{equation}
  \psi_1^{s_1s_2}(q_1)= \frac{1}{\sqrt{2}} (\delta_{s_{1},1}
  \delta_{s_{2},-1} - \delta_{s_{1},-1} \delta_{s_{2},1})
  \psi_1(q_1)\,,
\end{equation}
corresponding to
\begin{equation}
  \frac{1}{\sqrt{2}}
  \left(|\!\uparrow\,\rangle|\!\downarrow\,\rangle-
   |\!\downarrow\,\rangle |\!\uparrow\,\rangle \right) \otimes \psi_1
\end{equation}
in the standard $\sigma_{z}$ representation.

Until particle 1 reaches its magnet the Schr\"odinger evolution preserves
this form and $\hat{W}_\cond =\frac{1}{2}I\otimes
|\psi_{1}\rangle\langle\psi_{1}|$, where $\psi_{1}=\psi_{1}(t)$ obeys
Schr\"odinger's equation for particle 1.  Moreover, until particles 2
reaches its magnet (i.e., in the absence of a magnetic field acting on
particle 2), $\hat{H}_{2}$ involves no coupling between spin and
translational degrees of freedom, so that the form \eqref{eff2} is
preserved and $\hat{W}_\cond$ evolves unitarily according to
\eqref{WSchroedinger}, even after particle 1 has reached its magnet.  After
particle 1 has passed trough its magnet (but before particle 2 reaches its
magnet) $\hat{W}_\cond = \frac{1}{2} (\hat{W}_{\text{up}} +
\hat{W}_{\text{down}}) $, where $\hat{W}_{\text{up}}$, respectively
$\hat{W}_{\text{down}}$, corresponds to the pure state
$|\!\uparrow\,\rangle\otimes\psi_{\text{up}}$, respectively
$|\!\downarrow\,\rangle \otimes \psi_{\text{down}}$, the states to which $
|\!\uparrow\,\rangle \otimes\psi_{1}$, respectively
$|\!\downarrow\,\rangle\otimes\psi_{1}$, would evolve under the
Schr\"odinger evolution for particle 1.  After particle 2 reaches its
magnet, $\hat{W}_\cond $ no longer evolves unitarily (or even
autonomously).  Rather it collapses either to $\hat{W}_{\text{up}} $ or
$\hat{W}_{\text{down}}$ according to whether the initial configuration is
such that $Q_{2}$ ends up going down or up.

Throughout the course of the entire experiment $Q_{1}$ evolves
according to \eqref{WcondBohm}. (Note also that after particle 2 has
crossed its magnet, \eqref{eff2} is again approximately satisfied,
with $\Psi^\perp$ the wave packet that does not contain $Q_{2}$.)

We now turn to the relations between the various density matrices, and
discuss first the relation between $W_\cond$ and $W_\red$. $W_\red$ is
the average conditional density matrix, with the average taken with
respect to quantum equilibrium, i.e., over the ensemble in which
$Q=(Q_1,Q_2)$ is $|\Psi|^2$ distributed: 
\begin{equation}\label{wwred}
   {W_\red}^{s_1}_{s_1'}(q_1,q_1') = \int\limits_{\conf_1 \times \conf_2}
   dQ_1 \, dQ_2 \, |\Psi(Q_1,Q_2)|^2 \,
   {W_\cond}^{s_1}_{s_1'}(q_1,q_1')(Q_2).
\end{equation}
This relation makes clear that a system can have a conditional and a
reduced density matrix at the same time, the two being different
{}from each other: the conditional density matrix
of a system depends on the configuration $Q_{2}$ of its environment;
when this dependence is averaged out by taking the quantum equilibrium
expected value one obtains the reduced density matrix of the system.
(Note that for spin 0 \eqref{wwred} is the quantum equilibrium average
of $|\psi_{\cond}\rangle\langle \psi_{\cond}|$.)

Similarly, the combined (reduced statistical) density matrix is an
average of the conditional density matrix, with the average taken over
the ensemble in which $\Psi$ is $\mu$ distributed and, given $\Psi$,
$Q$ is $|\Psi|^2$ distributed:
\begin{equation}
   {W_\comb}^{s_1}_{s_1'}(q_1,q_1') = \int\limits_{\sphere}
   \mu(d\Psi) \int\limits_{\conf_1 \times \conf_2}
   dQ_1 \, dQ_2 \, |\Psi(Q_1,Q_2)|^2 \,
   {W_\cond}^{s_1}_{s_1'}(q_1,q_1')(Q_2,\Psi).
\end{equation}
Of course, $W_\stat$ can also be viewed as an average (of $|\psi
\rangle \langle \psi |$) over the ensemble with $\mu$-distributed
$\psi$, but this does not involve the conditional density matrix.

The fact that $W_\cond$ determines the Bohmian velocities according to
\eqref{WcondBohm} should be contrasted with the failure of such a
connection for $W_\stat$, $W_\red$, and $W_\comb$: If the wave
function $\psi$ of a system is random, the Bohmian velocities have to
be computed {}from the actual realization of $\psi$, and thus could
assume different values, corresponding to different $\psi$'s, even
when $Q$ is held fixed. Inserting, for example, $W_\stat$ in a formula
like \eqref{WcondBohm} or \eqref{WBohm} would yield, in contrast, an
average velocity at $Q$, averaged over the ensemble of different
$\psi$'s  (with the additional $Q$-dependent weight proportional to
$|\psi(Q)|^{2}$). This is what Bell referred to in the phrase we
quoted in the beginning, and what he elucidated in \cite{Belldensity}.
Similarly, since $W_\red$ is the average of the conditional density
matrix, over a certain ensemble, it leads to an average velocity (in
fact to the best guess at the velocity that one could make without
knowing $Q_2$). In contrast, $W_\cond$ depends on the actual value of
$Q_2$ and yields the true Bohmian velocity, as defined by \eqref{Bohm}
and the wave function of the universe.

The statistical analysis of Bohmian mechanics in \cite{qe} remains
valid when conditional wave functions are replaced by conditional
density matrices.

\section{Remarks}

\subsection{Conditional Density Matrix in Orthodox Quantum Mechanics}

In orthodox quantum mechanics, the definition \eqref{Wconddef} of the
conditional density matrix cannot be written down, for lack of a
configuration $Q_2$ that could be inserted into $\Psi$.  However,
orthodox quantum mechanics arguably maintains that macroscopic objects
can be viewed and treated classically, which presumably means that
there should exist something like a ``macroscopic configuration.'' In
case that $\Psi$ is such that the conditional density matrix does not
change much with the microscopic details of $Q_2$ (i.e., that it is
quite accurately determined by merely the macroscopic information
about $Q_2$), a conditional density matrix also makes sense in
orthodox quantum mechanics. In this case the conditional density
matrix of orthodox quantum mechanics would equal, within its accuracy,
the one of Bohmian mechanics. Another way of obtaining this density
matrix is to collapse the wave function (to the region of
configuration space having $q_2$ compatible with the actual
macroscopic configuration of $S_2$), and then to take the reduced
density matrix.

\subsection{Second Quantization}

In \cite{crea2}, we describe a construction that might be called the
``second quantization of a Markov process.'' Parallel to the ``second
quantization'' algorithm of forming a Fock space out of a given
1-particle Hilbert space and the free Hamiltonian on Fock space out of
a given 1-particle Hamiltonian, this construction builds a dynamics on
the configuration space of a variable number of particles out of a
given 1-particle dynamics. A key step in this construction is a
general procedure for forming the law of motion for $N$ particles,
given an \emph{arbitrary} 1-particle law. Interestingly, the
conditional density matrix is indispensable for this procedure (except
when wave functions are complex--valued).

A Bohm-type law of motion for one particle associates a velocity
vector field on $\RRR^3$ with every (smooth) 1-particle wave function.
We now regard this association abstractly as a given mapping, {}from
which we want to systematically construct the $N$-particle law that
provides the velocities of all particles {}from an $N$-particle wave
function and the positions of all particles. By inserting the
positions of all but one particle into the wave function, we get a
conditional object for one particle---for spin 0 a conditional wave
function, otherwise a conditional density matrix. Only if the
one-particle law associates with this conditional object a velocity
vector field on $\RRR^3$, can we insert the position of the remaining
particle into the vector field and get the particle's velocity.
 For $\text{spin} > 0$ we thus need more than what
we mentioned at the beginning of this paragraph: we need that the
one-particle law provide a velocity field \emph{for every density
   matrix}, as $W$-Bohmian mechanics does, and not merely for every
wave function.

\subsection{Empirical Consequences of $W$-Bohmian Mechanics}

One may wonder whether one can decide empirically between Bohmian
mechanics and $W$-Bohmian mechanics, or, in other words, whether one
can determine empirically in a universe governed by $W$-Bohmian
mechanics if the fundamental density matrix is pure \eqref{Wpsi}. The
question is delicate. We think that the answer is no, for the
following reason: compare a $W$-Bohmian universe with a Bohmian
universe with a random wave function such that the associated
statistical density matrix equals the fundamental density matrix of
the $W$-Bohmian universe. Since an empirical decision, if it can be
made at time $t_0$, would have to be based solely on the configuration
$Q_{t_0}$ at that time, and since the distribution of $Q_{t_0}$ is the
same in both situations, it seems that there cannot be a detectable
difference: A given $Q_{t_0}$ could as well have
arisen from an appropriate wave function from the random wave function
ensemble as from the corresponding fundamental density matrix.

What makes the question delicate, however, is, in part, the following:
we might not take seriously a theory involving a wave function of the
universe or a density matrix of the universe that is ``unreasonable''
or ``conspiratorial.''  Therefore, the question is connected to
questions such as what would count as a ``reasonable'' $W_\fund$, and
whether a statistical mixture mimicking a given ``reasonable''
$W_\fund$ might have to contain some ``unreasonable'' wave functions.

\subsection{Conditioning on Spatial Regions}\label{sec:region}

It is often desirable to define the subsystems $S_i$, $i=1,2$, as
encompassing all those particles which are presently located in the
regions $R_i \subseteq \RRR^3$, with $R_1 \cup R_2 = \RRR^3$ and $R_1
\cap R_2 = \emptyset$. To condition on the configuration $Q_2$ of
$S_2$ then means to condition on the configuration in the region
$R_2$. We describe below what appears to be the most convenient way to
carry out such a conditioning on a spatial region.  One might suspect
that conditioning on a spatial region is a very complicated story.
But, in fact, it could not be simpler.

Since the number of particles in the region $R_i$ can vary over time,
it is helpful to consider right {}from the start a configuration space
of a variable number of particles. We consider the space
\begin{equation}\label{Gamma}
   \Gamma (\RRR^3) := \bigcup_{n=0}^\infty \RRR^{3n}/S_n
\end{equation}
where $S_n$ denotes the group of permutations of $n$ objects, which
acts on $\RRR^{3n}$ by permuting the particle labels. A configuration
{}from $\Gamma(\RRR^3)$ represents any number of identical (unlabeled)
particles. For a discussion of this space, see \cite{crea2}.

We can extend the definition of $\Gamma$ to arbitrary sets $R$,
\begin{equation}
   \Gamma(R) :=  \bigcup_{n=0}^\infty R^{n}/S_n.
\end{equation}
When $R \subset \RRR^3$, $\Gamma(R)$ can be viewed as a subset of
$\Gamma(\RRR^3)$, containing those configurations for which all
particles are located in $R$. Now observe that, when $R_1$ and $R_2$
are disjoint sets, then
\begin{equation}\label{GammaR}
   \Gamma(R_1 \cup R_2) = \Gamma(R_1) \times \Gamma(R_2)\,.
\end{equation}
This property is helpful, as it tells us that the definition of the
subsystems $S_i$ in terms of spatial regions $R_i$ leads to a
Cartesian product decomposition $\conf = \conf_1 \times \conf_2$ of
configuration space, and thus allows us to use, without change, all of
our considerations on conditional density matrices, which assumed such
a decomposition.

\section{Conclusions}

We have introduced the notion of conditional density matrix in Bohmian
mechanics, and contrasted it, on the one hand, with the notion of
conditional wave function, and on the other hand, with various other
notions of density matrices. In contrast to the statistical, reduced,
or combined (reduced statistical) density matrix, the conditional
density matrix possesses direct significance for the particle
velocities.

The fact that with the same system can be associated several density
matrices brings into sharp focus that the \emph{meaning} of a density
matrix is not a priori; instead, various meanings are conceivable.
Ultimately, the meaning of a density matrix arises {}from its
relevance to the primitive objects, such as particle world lines, that
the theory is about.  In Bohmian mechanics, the various types of
density matrices that we have considered are all relevant to the
particles, but in very different ways.

\end{document}